\documentclass{trbunofficial-ad}

\usepackage{graphicx}
\usepackage{amsmath}
\usepackage{amssymb}
\usepackage{array}
\usepackage{bbm}
\usepackage[colorlinks=true, linkcolor=blue, citecolor=blue, urlcolor=blue]{hyperref}

\graphicspath{{images/}{../}{./}}

\makeatletter
\long\def\@makecaption#1#2{%
  \vskip\abovecaptionskip
  {\raggedright\bfseries\boldmath #1\;#2\par}%
  \vskip\belowcaptionskip}
\makeatother

\usepackage[backend=bibtex, style=authoryear, sorting=nyt, natbib=true,
            giveninits=false, doi=true, url=false, isbn=false, eprint=false]{biblatex}
\usepackage{zref-totpages}
\DeclareFieldFormat{title}{\mkbibquote{#1}}
\DeclareFieldFormat[article]{title}{\mkbibquote{#1}}
\AtBeginBibliography{\emergencystretch=3em\relax}
\AuthorHeaders{R. Bello, A. Mukwaya, G. Comert, V. Vaidyan, V. Bendigeri, A. Dontoh, J. Sahoo, and J. Mwakalonge}
\title{NS3Learn: Transferring 5G NR Mode-2 Reception Realism from ns-3 to the Veins/SUMO Stack for Connected-Vehicle Safety Assessment}

\author{%
  \textbf{Rasheed Bello*}\\
  South Carolina State University\\
  Orangeburg, SC 29117, USA\\
  Email: rbello@scsu.edu\\
  \hfill\break
  \textbf{Arthur Mukwaya}\\
  South Carolina State University\\
  Orangeburg, SC 29117, USA\\
  Email: amukwaya@scsu.edu\\
  \hfill\break
  \textbf{Gurcan Comert}\\
  North Carolina A\&T State University\\
  Greensboro, NC 27411, USA\\
  Email: gcomert@ncat.edu\\
  \hfill\break
  \textbf{Varghese Vaidyan}\\
  Dakota State University\\
  Madison, SD 57042, USA\\
  Email: varghese.vaidyan@dsu.edu\\
  \hfill\break
  \textbf{Vijay Bendigeri}\\
  Independent Researcher\\
  Sacramento, CA, USA\\
  Email: vijaybendigeri@gmail.com\\
  \hfill\break
  \textbf{Anthony Dontoh}\\
  South Carolina State University\\
  Email: adontoh@scsu.edu\\
  \hfill\break
  \textbf{Sahoo Jagruti}\\
  South Carolina State University\\
  Email: jsahoo@scsu.edu\\
  \hfill\break
  \textbf{Judith Mwakalonge}\\
  South Carolina State University\\
  Email: jmwakalo@scsu.edu\\
  \hfill
}

\begin{document}

{\fontsize{12}{14.4}\selectfont
\maketitle}
\thispagestyle{empty}
\newpage

\pagestyle{main}

{\fontsize{12}{14.4}\selectfont\section*{Abstract}}

\noindent\textbf{Objectives}\hspace{0.6em}
Connected-vehicle safety studies estimate deployment benefits with coupled
traffic and network simulation, and their conclusions depend on which messages arrive. Common channel models resolve radio
propagation but omit the competition for radio resources that governs 5G New
Radio sidelink Mode-2 and related sidelink modes, so they report near-perfect
delivery in dense traffic where real deployments lose most messages. This work adds that behavior
without building a protocol implementation.\\

\noindent\textbf{Methods}\hspace{0.6em}
We labeled 10.5 million reception outcomes from the physical-layer and
scheduler traces of ns-3 5G-LENA, a reference implementation calibrated on
Third Generation Partnership Project scenarios and driven by Simulation of Urban
Mobility trajectories. From these labels we fitted NS3Learn, a closed-form model
whose stages represent half-duplex loss, scheduling collisions, receiver capture,
and decoding. Experiments spanned two signalized urban networks, six penetration
levels from 1 to 100 percent, and five seeds per condition, three on the
corridor, scoring NS3Learn against the unmodified channel and a combined
reference from two published models.\\

\noindent\textbf{Findings}\hspace{0.6em}
NS3Learn tracked ns-3 5G-LENA to a mean absolute deviation of 0.06 in
per-instant delivery, an in-sample figure, while the two alternatives deviated by 0.44 and 0.55 against per-seed variation of 0.03. Coefficients fitted at one intersection carried to a
structurally different junction at about 20 percent more error, and changing only the
communication model reversed the direction of the simulated speed trend and
more than doubled hard braking.\\

\noindent\textbf{Novelty}\hspace{0.6em}
Reception realism transfers between simulators by distillation rather than
reimplementation or analytical derivation. Each stage represents one named
mechanism, so every published coefficient traces to the physical effect it
encodes.\\

\noindent\textbf{Practical Applications}\hspace{0.6em}
Agencies and researchers can keep the pipelines they already operate and still
represent the losses that dominate dense traffic at signalized intersections
and in sidelink denial-of-service studies.
Retargeting to another radio configuration means repeating the offline fitting,
not modifying simulation code.\\

\hfill\break%
\noindent\textit{Keywords}: connected and automated vehicles, 5G NR sidelink, V2X simulation, medium-access contention, knowledge distillation, traffic safety
\newpage

\section{Introduction}
Connected-vehicle safety applications depend on the timely exchange of basic safety messages, and simulation quantifies their benefits long before field deployment. Those simulated benefits (directly or indirectly) inform deployment plans, infrastructure investment, and the safety cases regulators read. The dominant evaluation platform couples the OMNeT++ discrete-event network simulator, the Veins vehicular framework, and the SUMO microscopic traffic simulator, so that road traffic and radio traffic influence each other within a single experiment \citep{sommer_bidirectionally_2011,lopez_microscopic_2018}. A safety conclusion drawn on this stack, a change in time-to-collision or in mean speed, is only as trustworthy as the model deciding which messages arrive and when.

The stack matured around dedicated short-range communications based on IEEE 802.11p, and its channel models target that setting. They resolve path loss, fading, and shadowing, but not the way vehicles compete for radio resources in 5G NR PC5 sidelink Mode-2. Vehicles in Mode-2 pick their own transmission resources through sensing-based semi-persistent scheduling (SB-SPS), with no base station to allocate them. Three mechanisms then govern delivery more strongly than propagation: collisions, when two vehicles independently pick the same resource; half-duplex periods, when a vehicle transmits and therefore cannot hear a neighbor transmitting at the same moment; and capture, when a receiver decodes the stronger of two overlapping signals \citep{garcia_tutorial_2021,ali_3gpp_2021}. All three sharpen as vehicles accumulate, including where propagation conditions are benign. A channel model that resolves only propagation therefore reports messages arriving that a real deployment would lose, and it does so most severely at the dense intersections where cooperative safety applications are supposed to earn their value.

The stack carries no mature model of NR PC5 Mode-2. Simu5G brings 5G NR to OMNeT++ but models the cellular Uu interface rather than the PC5 sidelink \citep{nardini_simu5gomnet_2020}. The one open OMNeT++ sidelink model, OpenCV2X, implements LTE-V2X Mode-4 and has not tracked NR \citep{mccarthy_opencv2x_2021}. The standards-compliant NR Mode-2 implementation lives in ns-3 instead, as an extension of the 3GPP-calibrated 5G-LENA simulator \citep{ali_3gpp_2021,patriciello_e2e_2019,koutlia_calibration_2022}. Researchers wanting NR Mode-2 realism therefore choose between abandoning the SUMO-coupled stack their safety pipeline is built on and accepting a communication model that omits the dominant loss mechanisms.

We explore a third option. Rather than port a protocol stack, we distill the Mode-2 reception behavior of ns-3 5G-LENA into a compact model that runs inside Veins. We call it NS3Learn. SUMO trajectories drive ns-3 5G-LENA across vehicle densities, and we label reception outcomes from its physical- and MAC-layer traces. Those labels train a cascade of stages that correspond one-to-one to half-duplex loss, SB-SPS collision, capture, and decoding, so every fitted number belongs to a named mechanism and can be checked against it. Every stage is a logistic function of vehicle density or signal-to-interference-plus-noise ratio, which keeps the composed model closed-form, readable, and cheap enough to evaluate for every message in an event-driven simulation. The cost of fidelity moves offline, into one labeling and fitting campaign, and the run time gains no protocol module to maintain.

We score NS3Learn against ns-3 5G-LENA and against the two alternatives a practitioner would otherwise use. The first is the stack base channel (naive DSRC Implementation), left untouched. The second pairs two published NR Mode-2 models that partition reception between them, one governing medium access under SB-SPS \citep{cao_toward_2026} and one governing propagation \citep{rehman_impact_2023}. Their components also serve as contention-only and propagation-only ablations. Neither reference represents capture, and neither is fitted to ns-3 5G-LENA. The comparison centers on the per-instant, in-range packet delivery ratio, the quantity a downstream metric depends on, rather than a time-averaged ratio that dilutes congestion effects under transient neighbor topologies \citep{toghi_spatio-temporal_2019}. A reception model matters where it changes a conclusion, so we follow each representation through to the driving behavior it induces, first under ordinary traffic and then under an adversarial flooding load that stresses the SB-SPS mechanism NS3Learn models \citep{twardokus_toward_2023}.

This paper makes four contributions. A distillation method transfers NR PC5 Mode-2 realism, across propagation and medium-access contention, from 3GPP-calibrated ns-3 5G-LENA into the Veins, OMNeT++, and SUMO stack, and leaves no protocol module behind to maintain. The reception cascade it produces is closed-form and evaluated per message, and its stages map one-to-one to half-duplex, collision, capture, and decoding. Its coefficients appear here in full, and those fitted at one signalized junction carry to a structurally distinct second junction with no re-estimation, at about 20 percent more error. A combined analytical reference for NR Mode-2 delivery joins a medium-access model to a propagation model at a boundary that charges interference once, runs alongside the unmodified stack, and resolves into single-axis ablations that attribute degradation to contention. Finally, the choice of reception model reaches the quantities a safety study reports: it reverses the direction of the speed trend, more than doubling hard braking under ordinary traffic, and registers an adversarial flood that neither reference represents at any flood rate.

\section{Background and Related Work}

\subsection{V2X Simulation for Safety in the Coupled Stack}
Bidirectional coupling of a network simulator and a traffic microsimulator is the established basis for V2X evaluation, since offline mobility traces cannot capture the feedback from communication back onto driving \citep{sommer_bidirectionally_2011,lopez_microscopic_2018}. Extensions toward cellular V2X have gone in two directions, and the sidelink falls between them. Simu5G adds a maintained 5G NR data plane to OMNeT++, scoped to the cellular Uu interface rather than the PC5 sidelink \citep{nardini_simu5gomnet_2020}. OpenCV2X supplies the sidelink but models LTE-V2X Mode-4, and shows the LTE sensing scheduler degrading under the aperiodic patterns typical of cooperative awareness traffic \citep{mccarthy_opencv2x_2021}. NR Mode-2 evaluation has largely taken place outside OMNeT++, on ns-3 or dedicated MATLAB tooling \citep{ali_3gpp_2021,todisco_performance_2021}.

\subsection{What Governs NR Mode-2 Delivery, and How It Has Been Modeled}
Vehicles sense the channel, exclude recently reserved resources, and select from the remainder, holding a reservation for a randomized number of transmissions before reselecting \citep{garcia_tutorial_2021,ali_3gpp_2021}. An analytical decomposition for the LTE-V2X predecessor attributes packet loss to four causes: half-duplex transmission, received power below the sensing threshold, propagation, and packet collisions \citep{gonzalez-martin_analytical_2019}. That four-way structure recurs across the literature and maps onto the stages of the distilled cascade below. For NR Mode-2, the analytical treatment splits along the same physical-versus-access boundary. One line derives the steady-state SB-SPS collision probability and resulting reception ratio in closed form, covering collisions from simultaneous reselection and from persistent reservation keeping, and validates the result against ns-3 5G-LENA \citep{cao_toward_2026}. A second line models multiple-access interference at the physical layer, approximating SINR statistics under composite fading to obtain the packet-success probability \citep{rehman_impact_2023}. Later work adds the standardized re-evaluation step \citep{zhu_analytical_2025} and compares semi-persistent against dynamic scheduling \citep{lusvarghi_comparative_2023}. Each result holds within its own stochastic-geometry or fixed-traffic assumptions, and each covers one slice of the mechanism rather than the joint behavior. The two contention-aware members form the analytical baselines here.

\subsection{Physical-Layer Abstraction Versus Learned Realism}
A body of work speeds up system-level simulation by abstracting the physical layer rather than simulating it symbol by symbol. Link-to-system mapping based on the exponential effective signal-to-interference-plus-noise ratio compresses per-subcarrier conditions into an effective value and reads block error rate from precomputed tables, the method underlying the physical-layer realism of 5G-LENA itself \citep{lagen_new_2020}. Later work specializes the technique to the NR sidelink, producing a transport-block error model for a single PSSCH link \citep{cao_efficient_2023}, and a related method for direct V2X reduces measured error-rate curves to one threshold plus an implementation-loss parameter, avoiding new link-level campaigns \citep{wu_methodology_2022}. Two limits separate these abstractions from a distilled model. Modeling one link at a time excludes the multi-vehicle contention that dominates Mode-2 delivery, and the NR sidelink abstraction is derived with interference set to zero \citep{cao_efficient_2023}. Each also stays inside the simulator that produced it. Our model learns the system-level delivery outcome with contention included, and it moves to a different simulator.

\subsection{Learned Surrogates and Context-Fitting}
A compact model fitted to an expensive simulator escapes the accuracy-versus-cost trade-off between analytical models and packet-level simulation \citep{matsumoto_neural_2026}. In the V2X setting, machine-learning models predict quality-of-service metrics such as packet delivery ratio and throughput from radio and context features, under validation protocols that guard against optimistic evaluation \citep{yazici_nr-v2x_2024}. The nearest work proposes a neural surrogate for autonomous-driving communications trained from a coupled ray-tracing, cellular, and traffic simulator, targeting cellular throughput and latency for on-vehicle inference rather than for injection back into a network simulator \citep{matsumoto_neural_2026}. Fitting realism to a deployment context has precedent: one study calibrates a Veins channel model against field-collected urban delivery data \citep{gammaa_improving_2025}. Our cascade shares that premise and differs in three respects. It targets NR PC5 Mode-2 per-message reception instead of cellular aggregates. Its fitting source is a controllable 3GPP-calibrated implementation, not a field campaign. It then runs in the Veins stack rather than in the engine that produced its labels.

\subsection{Communication Realism and Its Effect on Driving}
Crashes are rare, so connected-vehicle safety studies rely on surrogate proxies such as time-to-collision. A systematic review of these measures ties their validity for connected and automated vehicles to how accurately the simulation models vehicle behavior and communication \citep{wang_review_2021}. A vehicle's trajectory responds to which messages arrived, so every measure computed on that trajectory inherits the delivery model. The stakes are clearest under attack. Protocol-aware denial-of-service attacks exploit the predictability of SB-SPS resource selection, cutting a target vehicle's delivery ratio within seconds and degrading roadway operations \citep{twardokus_toward_2023}. Resource-starvation attacks on the same mechanism raise information age and safety risk while averaged delivery metrics fail to detect them \citep{ashik_analyzing_2026}. Studying how either translates into traffic outcomes inside the standard stack requires a channel model that represents SB-SPS.

\section{Methodology}
The method has four parts: the reference implementation that supplies the labels, the analytical baseline and the distilled cascade measured against it, the fitting and transfer protocol, and the evaluation metrics.

\subsection{The Reference Implementation}
The reference implementation for reception is the NR V2X Mode-2 sidelink implementation of ns-3 5G-LENA, a system-level NR simulator calibrated against 3GPP reference scenarios \citep{ali_3gpp_2021,patriciello_e2e_2019,koutlia_calibration_2022}. Vehicles transmit broadcast safety messages using SB-SPS autonomous resource selection with sensing enabled. Propagation follows the 3GPP TR 37.885 V2V-Urban model with vehicle-height antennas and non-line-of-sight-by-vehicle blockage \citep{3gpp_37885}, the channel appropriate to sidelink and distinct from the cellular street-canyon models that serve Uu links. Both simulators run the same traffic context: one road network, demand pattern, penetration level, and seed per repetition. The cascade reads transmitter-receiver distance and in-range density, so it responds to the geometry it is placed in, and a shared network and demand is what grounds the delivery comparison.

The two simulators play different roles. ns-3 5G-LENA resolves the radio and nothing else, replaying an exported trajectory file so that driving is prescribed and cannot respond to which messages arrive. Prescribed mobility is standard for system-level radio evaluation and costs nothing when delivery is the only quantity of interest, though it forecloses questions about how communication quality shapes traffic. Veins can close that loop, driving SUMO over TraCI so that a vehicle acts on the messages it receives. The distilled cascade combines the two: a Mode-2 reception model inside a stack where driving can answer to it.

The delivery and trajectory experiments differ in how far communication acts on driving. The delivery experiments disable both fail-safe responses and the message-driven controller switch, so connected vehicles hold the cooperative car-following model throughout and none of them changes that model on the strength of a received message. Signal-phase messages still inform the approach to a stop line, so the coupling is reduced rather than removed, and trajectories at a given seed stay close without being identical across treatments. The trajectory experiments enable the fail-safe, since a reception model can change driving only if driving answers to it, and there seeds of one configuration span a factor of two in total receptions. We report means over repetitions throughout.

\begin{figure}[!ht]
  \centering
  \includegraphics[width=1\textwidth]{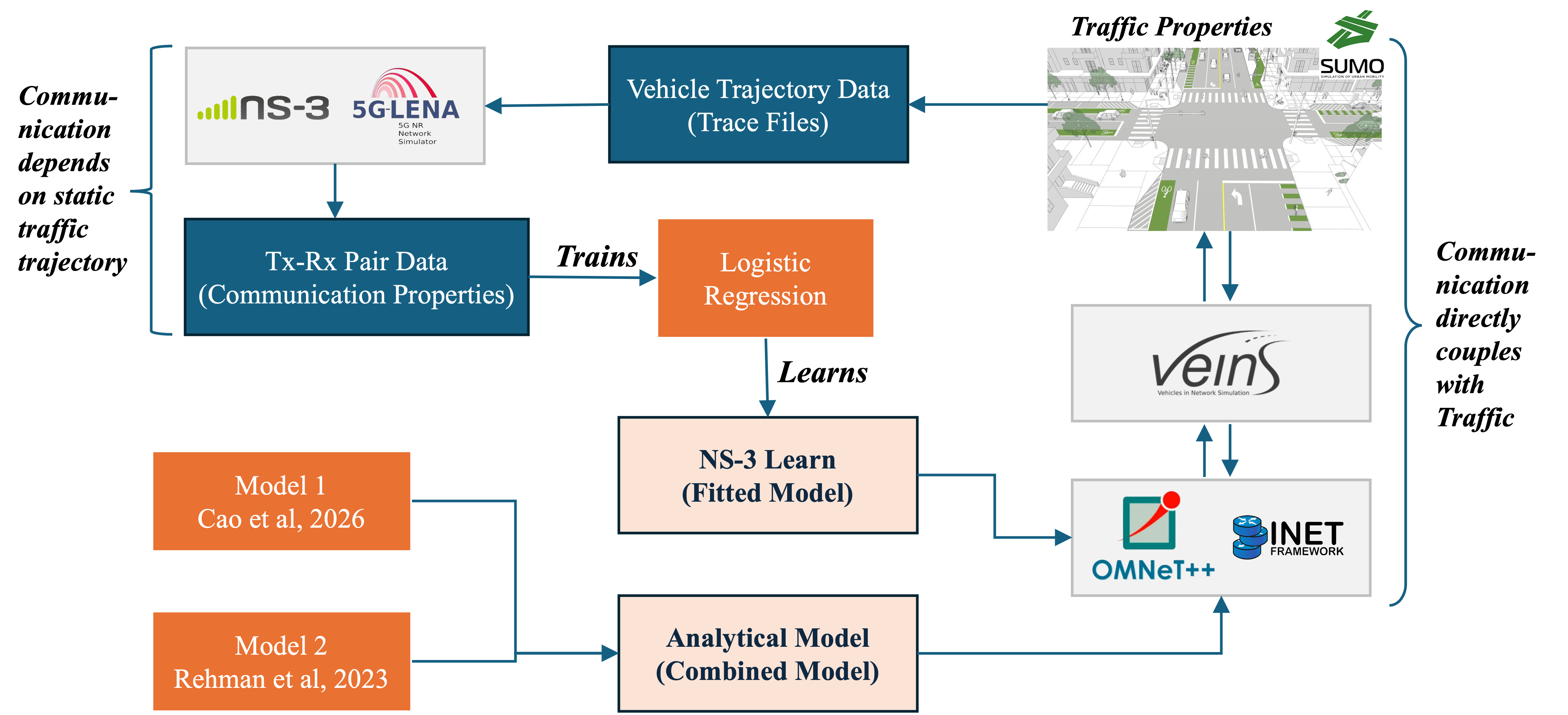}
  \caption{Distillation pipeline. ns-3 5G-LENA replays a fixed SUMO trajectory file and supplies the fitting labels. The fitted model then runs inside OMNeT++/INET beside the combined analytical reference, where Veins and SUMO exchange state each step and delivery feeds back into driving.}\label{fig:pipeline}
\end{figure}

Hybrid automatic repeat request is active, and a reservation may carry up to three transmissions of a transport block. The reference implementation records $2.5$ on average. A message counts as delivered once any of its transmissions decodes, so the fitted labels absorb the effect and it does not appear as a separate stage. Figure~\ref{fig:pipeline} traces the pipeline: SUMO mobility feeds ns-3 5G-LENA, whose physical- and MAC-layer traces reduce to per-stage labels, which we fit to a cascade and load into Veins for scenario studies.

\subsection{Analytical Baseline}
For NR PC5 Mode-2, no published analytical model matches the completeness the four-cause decomposition achieved for LTE Mode-4 \citep{gonzalez-martin_analytical_2019}. The closest attempt combines persistence and propagation to study the Age of Information, through a mean-field system-wide success probability rather than a per-link reception model. It also treats overlapping signals as certain loss, leaving capture unmodelled \citep{rolich_understanding_2024}. We therefore build a combined reference from two published Mode-2 models, each supplying the mechanism it was derived to represent, and treat its factors as contention-only and propagation-only ablations. This reference is a baseline rather than a contribution, so we state only what our implementation evaluates and refer to the source papers for the derivations.

\textit{Medium access.} The first model treats delivery as the outcome of SB-SPS resource collisions with an error-free physical layer \citep{cao_toward_2026}. A vehicle holds a resource for a reselection counter $R_c\sim U(5,15)$ and reselects with probability $1-p_k$ when it expires. We count candidate resources over the window a transmitter selects from. 3GPP TS 38.214 §8.1.4 confines selection to $[n+T_1,\,n+T_2]$, so that window rather than the reservation period sets the count: $N_r=W\rho N_{sc}$ for a window of $W$ slots, $N_{sc}$ subchannels per slot, and a sidelink slot-pool fraction $\rho$. Our implementation counts the window as $T_2$ slots, giving $W=33$, with $\rho=9/12$ from the pool bitmap and $N_{sc}=5$, hence $N_r\approx124$. Counting the span as $T_2-T_1$ instead would give $N_r\approx116$, which moves no result reported here. Cao counts sidelink slots alone on an all-sidelink pool where $\rho=1$; the $\rho$ factor generalizes the count to a mixed pool. The stationary reselection probability is $\pi_0=1/11$ per interval, and ns-3 5G-LENA disables resource keeping, so $p_k=0$ throughout and the general expression collapses to
\hfill\break
\begin{linenomath}
\begin{equation}
P_{\mathrm{COL}}=1-\left(1-\frac{2\pi_0}{\overline{N_a}}\right)^{N_{UE}},
\label{eq:pcol}
\end{equation}
\end{linenomath}
\hfill\break
which is what our implementation evaluates. The mean count of unoccupied resources $\overline{N_a}$ depends on $P_{\mathrm{COL}}$ in turn, through the share of collided resources carrying $\overline{N_c}=2$ packets in the under-saturated regime, so the pair is solved jointly. Half-duplex loss is the slot fraction $P_{\mathrm{HD}}=t_s/T_{RRI}$, fixed at $0.01$ and independent of density. This model represents contention faithfully but assumes error-free decoding absent a collision, carrying no distance dependence beyond the range cutoff.

\textit{Propagation and decode.} The second model treats delivery through the physical layer \citep{rehman_impact_2023}. Received power follows an inverse-power law with Nakagami fading and log-normal shadowing, and the signal-to-interference-plus-noise ratio divides it by the aggregate co-resource interference plus noise. That aggregate has no closed-form density, so moment matching replaces it with a single log-normal. Integrating the block-error rate against the resulting density $p_{\hat\gamma}(x)$ gives the packet-success probability,
\hfill\break
\begin{linenomath}
\begin{equation}
p(d)=1-\int_0^{\infty}\mathrm{PER}(x)\,p_{\hat\gamma}(x)\,dx,
\label{eq:psucc}
\end{equation}
\end{linenomath}
\hfill\break
with $\mathrm{PER}(x)$ read from block-error-rate curves. We depart from the source in one respect: in place of its inverse-power law we substitute the TR 37.885 V2V-Urban path loss that ns-3 5G-LENA uses, retaining that model's line-of-sight probability, vehicle-blockage excess, and log-normal shadowing. Giving the reference and ns-3 5G-LENA the same channel isolates the comparison on medium access, which a propagation mismatch would otherwise obscure.

\textit{Combined reference.} The two models occupy disjoint parts of the pipeline and join at their boundary: the first governs whether the medium impairs a transmission, through $P_{\mathrm{COL}}(n)$ and $P_{\mathrm{HD}}$, and the second governs whether a clean transmission decodes at distance. Counting interference twice is the one hazard, since collisions are already resolved as MAC events. The decode term is therefore taken in its noise-limited form, where the SINR reduces to the SNR defined for the no-interferer case \citep{rehman_impact_2023}, giving
\hfill\break
\begin{linenomath}
\begin{equation}
\mathrm{PDR}_{\mathrm{M3}}(d,n)=(1-P_{\mathrm{HD}})\,\bigl(1-P_{\mathrm{COL}}(n)\bigr)\,g\!\bigl(\mathrm{SNR}(d)\bigr),
\label{eq:m3}
\end{equation}
\end{linenomath}
\hfill\break
where $g$ is Equation~\eqref{eq:psucc} evaluated with an empty interferer set. We write M3 for this combination of the two source models, and use that label in the figures and tables. Density drives the first two factors, distance the third, and capture appears in none of them. Equation~\eqref{eq:m3} is the analytical counterpart of the distilled cascade, differing in that it omits capture and that its factors are derived under stochastic assumptions rather than fitted.

Three scope conditions apply. The collision model is validated against ns-3 for transmitting-vehicle counts $N_{UE}\in[20,200]$ over two subchannels, whereas our scenario spans in-range counts from roughly eight to about sixty across five subchannels, so the sparsest condition extrapolates below the validated range. That model also assumes an under-saturated channel and deviates near saturation \citep{cao_toward_2026}. The decode term reads a block-error-rate curve produced in a separate 5G Toolbox link simulation at 16-QAM with code rate $553/1024$, index 14 of the NR table and the setting ns-3 5G-LENA transmits at; no quantity derived from ns-3 output enters that curve. Reference and ns-3 5G-LENA therefore share channel and link configuration and differ only on medium access. Capture is absent by construction and no term is calibrated, so the deviations reported below follow from these stated choices rather than from tuning.

\subsection{The NS-3-Distilled Reception Cascade}
We model reception of a single message as a two-stage pipeline: a radio-frequency cascade distilled from ns-3, followed by an application-layer processing limit at the receiving on-board unit. The cascade decomposes delivery into physical mechanisms and represents each as a logistic function, $\sigma(z)=1/(1+e^{-z})$, of transmitter-receiver distance $d$, in-range transmitter count $n$, or post-path-loss signal-to-interference-plus-noise ratio $\gamma$. The decomposition mirrors the four-cause structure established for sidelink delivery \citep{gonzalez-martin_analytical_2019}, with capture treated explicitly so that resource collisions do not translate one-for-one into loss.

The noise-limited signal-to-interference-plus-noise ratio at distance $d$ follows from the link budget,
\begin{linenomath}
\begin{equation}
\gamma(d)=P_{\mathrm{tx}}-\mathrm{PL}(d)-N_0,
\label{eq:sinr}
\end{equation}
\end{linenomath}
\hfill\break
where $P_{\mathrm{tx}}$ is transmit power, $N_0$ is the noise floor, and $\mathrm{PL}(d)$ is the TR 37.885 V2V-Urban path loss with per-link line-of-sight state and first-order shadow correlation \citep{3gpp_37885}.

A receiver loses a message while it is itself transmitting. This half-duplex loss grows with the number of concurrent transmitters, and we model it as a logistic function of density,
\hfill\break
\begin{linenomath}
\begin{equation}
h(n)=\sigma\!\bigl(\eta_0+\eta_1 n+\eta_2 n^2\bigr).
\label{eq:hd}
\end{equation}
\end{linenomath}
\hfill\break
Independent SB-SPS selections collide when two transmitters reserve overlapping resources. The baseline collision probability, in the absence of any adversary, is likewise logistic in density,
\hfill\break
\begin{linenomath}
\begin{equation}
c_0(n)=\sigma\!\bigl(\kappa_0+\kappa_1 n+\kappa_2 n^2\bigr).
\label{eq:coll0}
\end{equation}
\end{linenomath}
\hfill\break
An adversarial flooder injecting at rate $r$ raises the collision level. Each in-range active flooder $a$ contributes a bounded factor $q_a$, and survival multiplies over flooders, so an empty adversary set leaves the channel at its baseline,
\hfill\break
\begin{linenomath}
\begin{align}
q_a(r,n) &= \sigma\!\bigl(\omega_0+\omega_1\log_{10} r+\omega_2\log_{10}(n{+}1)+\omega_3\log_{10} r\,\log_{10}(n{+}1)\bigr),
\label{eq:qa}\\
c(n) &= 1-\bigl(1-c_0(n)\bigr)\!\!\prod_{a\in\mathcal{A}}\!\bigl(1-q_a(r_a,n)\bigr).
\label{eq:coll}
\end{align}
\end{linenomath}
\hfill\break
The transport block decodes with a probability that rises with signal-to-interference-plus-noise ratio and falls with density, whether or not a collision occurred. We represent capture with two decode branches, one for the collision-free case and one for the collided case, each a logistic surface in $\gamma$ and $n$, \hfill\break
\begin{linenomath}
\begin{equation}
g^{(b)}(\gamma,n)=\sigma\!\bigl(\theta^{(b)}_0+\theta^{(b)}_1\gamma+\theta^{(b)}_2\gamma^2+\theta^{(b)}_3 n+\theta^{(b)}_4\,\gamma n\bigr),
\qquad b\in\{\mathrm{nc},\mathrm{cc}\}.
\label{eq:decode}
\end{equation}
\end{linenomath}
\hfill\break
The collision branch $g^{(\mathrm{cc})}$ carries the interferer penalty implicitly, so a near transmitter survives a collision that a far transmitter does not. That asymmetry is the capture effect.

\begin{figure}[!ht]
  \centering
  \includegraphics[width=0.9\textwidth]{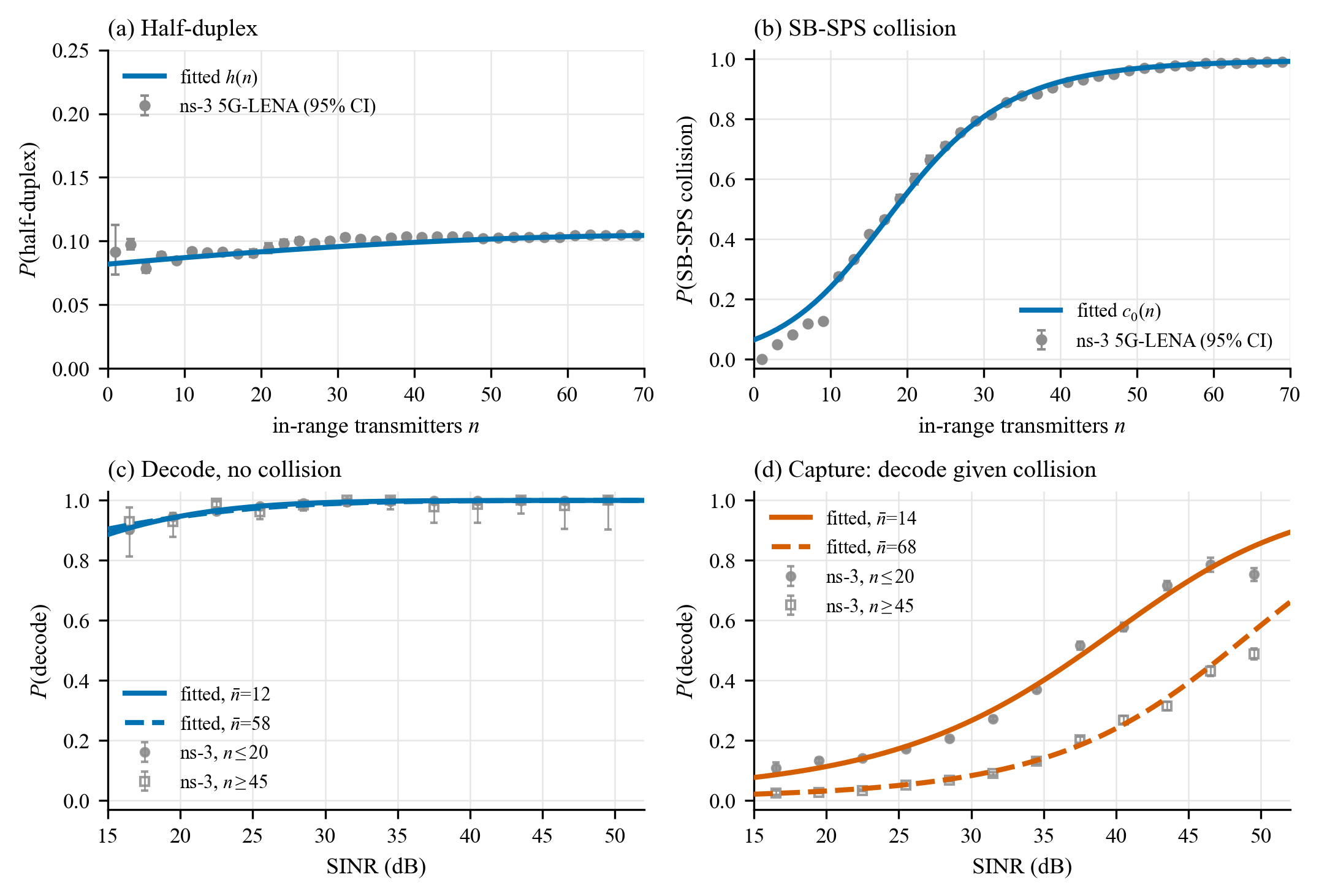}
  \caption{Each cascade stage against the ns-3 5G-LENA labels it was fitted to; markers are empirical proportions with 95\,\% Wilson intervals, curves the deployed coefficients evaluated at each band's mean density. The gap between (c) and (d), narrowing as neighbors accumulate, is the capture effect.}\label{fig:stages}
\end{figure}

Taking the expectation over the collision draw gives the radio-frequency delivery probability,
\hfill\break
\begin{linenomath}
\begin{equation}
p_{\mathrm{rf}}(d,n)=\bigl(1-h(n)\bigr)\Bigl[\bigl(1-c(n)\bigr)\,g^{(\mathrm{nc})}(\gamma,n)+c(n)\,g^{(\mathrm{cc})}(\gamma,n)\Bigr].
\label{eq:prf}
\end{equation}
\end{linenomath}
\hfill\break
Traffic exceeding the receiver's processing budget drops at the application layer. With sliding-window load $L$ over window $W$ and capacity $C=C_{\mathrm{obu}}W$, the overload drop is
\hfill\break
\begin{linenomath}
\begin{equation}
p_{\mathrm{obu}}=\min\!\Bigl(1,\; s\cdot\max\!\bigl(0,\tfrac{L-C}{C}\bigr)\Bigr),
\label{eq:obu}
\end{equation}
\end{linenomath}
\hfill\break
and the per-message delivery probability combines the two stages,
\hfill\break
\begin{linenomath}
\begin{equation}
\mathrm{PDR}(d,n)=p_{\mathrm{rf}}(d,n)\,\bigl(1-p_{\mathrm{obu}}\bigr).
\label{eq:pdr}
\end{equation}
\end{linenomath}
\hfill\break
Table~\ref{tab:coeffs} lists the fitted coefficients of Equations~\eqref{eq:hd}--\eqref{eq:obu}; the radio constants that Equation~\eqref{eq:sinr} reads are scenario-fixed and appear in Table~\ref{tab:scenario}. All fitted values come from ns-3 5G-LENA by the protocol described below. The simulation reads them from a plain-text file when a run starts, so moving the model to a new radio configuration means supplying new numbers rather than editing and rebuilding the simulator. Figure~\ref{fig:stages} shows the fitted stages against the ns-3 5G-LENA labels across the density and signal-to-interference-plus-noise-ratio range.

\begin{table}[!ht]
\caption{Fitted Coefficients of the Distilled Reception Cascade}\label{tab:coeffs}
\begin{center}
\footnotesize
\begin{tabular}{|l|l|l|}
\hline
\textbf{Symbol} & \textbf{Meaning (equation)} & \textbf{Value}\\
\hline
$\eta_0,\eta_1,\eta_2$ & Half-duplex $h(n)$, Eq.~\eqref{eq:hd} & $-2.41737,\ 7.05\!\times\!10^{-3},\ -4.61\!\times\!10^{-5}$ \\
\hline
$\kappa_0,\kappa_1,\kappa_2$ & Baseline collision $c_0(n)$, Eq.~\eqref{eq:coll0} & $-2.66873,\ 0.15898,\ -7.39\!\times\!10^{-4}$ \\
\hline
$\omega_0,\omega_1,\omega_2,\omega_3$ & Flood factor $q_a$, Eq.~\eqref{eq:qa} & $-0.037,\ -0.086,\ -0.398,\ 0.268$ \\
\hline
$\theta^{(\mathrm{nc})}_{0\text{--}4}$ & Decode, no collision, Eq.~\eqref{eq:decode} & $0.00686,\ 0.09830,\ 0.00231,\ 0.01918,\ -9.92\!\times\!10^{-4}$ \\
\hline
$\theta^{(\mathrm{cc})}_{0\text{--}4}$ & Decode, collision (capture), Eq.~\eqref{eq:decode} & $-3.06368,\ 0.04412,\ 1.22\!\times\!10^{-3},\ -0.02376,\ -6.80\!\times\!10^{-5}$ \\
\hline
\multicolumn{3}{|l|}{\textit{Assumed, not fitted: the reference implementation bounds no receiver processing}}\\
\hline
$C_{\mathrm{obu}},\,W,\,s$ & OBU capacity, window, drop slope, Eq.~\eqref{eq:obu} & 600 pkt/s,\ \ 1 s,\ \ 2.0 \\
\hline
\end{tabular}
\end{center}
\end{table}

\subsection{Fitting and Transfer Protocol}
We fit each stage to the signal that isolates it. Half-duplex and collision labels come from the MAC schedule and the concurrent-transmission summary, both available across the full density range. Decode labels come from the per-reception physical-layer trace, collected at 25, 50, and 100 percent penetration, since sampling it at high penetration exposes the density dependence of decoding. Capture weakens as neighbors accumulate, which motivates the density terms $\theta_3$ and $\theta_4$ in Equation~\eqref{eq:decode} and separates the two branches by more at low density than at high.

Density must enter per reception rather than per run. Labeling each reception with the run-mean neighbor count collapses the trace onto a few density levels and overfits them, and the fitted surface then transfers poorly. Computing the receiver's in-range neighbor count at the instant of each reception, matching how the run time evaluates the model, removes that bias and improves agreement across transmitter-receiver pairs from $0.085$ to $0.072$ mean absolute deviation.

We fit each logistic stage by $L_2$-penalized logistic regression and validate the composed cascade on the per-instant in-range delivery ratio. The penalty matters for reproduction: the decode surfaces carry collinear terms in $\gamma$, $\gamma^2$, and $\gamma n$, and an unpenalized refit returns materially different coefficients for the non-collision branch. Table~\ref{tab:coeffs} reports the penalized estimates, which are the values the run time loads. Generalization follows from fitting on one geometry and validating on another. The coefficients in Table~\ref{tab:coeffs} are estimated at a single signalized intersection, then frozen and applied unchanged to a two-intersection urban corridor of comparable footprint. We also run ns-3 5G-LENA on that corridor to supply an out-of-sample reference. Both layouts concentrate neighbors around controlled junctions, so the transfer probes sensitivity to layout, scale, and signal timing rather than any contrast between clustered and dispersed traffic.

\subsection{Evaluation Metrics}
Delivery and its loss attribution measure the channel. Mean speed and hard braking measure what a difference in delivery does to driving.

The primary communication metric is the per-instant, in-range packet delivery ratio, evaluated per link and aggregated over the messages $\mathcal{M}$ transmitted while their intended receivers are within communication range in the analysis window,
\hfill\break
\begin{linenomath}
\begin{equation}
\overline{\mathrm{PDR}}=\frac{1}{|\mathcal{M}|}\sum_{m\in\mathcal{M}}\mathbbm{1}\{m\ \text{delivered}\},\qquad
\Pr[m\ \text{delivered}]=\mathrm{PDR}(d_m,n_m).
\label{eq:prr}
\end{equation}
\end{linenomath}
\hfill\break
Evaluating delivery per instant and per link, rather than averaging over a window during which the set of in-range neighbors changes, keeps the metric sensitive to congestion \citep{toghi_spatio-temporal_2019}. The second metric partitions the losses: we charge every discarded message to the stage that discarded it, so one run yields both the delivery ratio and its attribution across half-duplex, contention and decode, and receiver overload. Attribution distinguishes a model reaching the right delivery level from one reaching it through the wrong mechanism. Attack potency is the delivery loss attributable to the flooder,
\hfill\break
\begin{linenomath}
\begin{equation}
\Delta\overline{\mathrm{PDR}}=\overline{\mathrm{PDR}}^{\,\text{baseline}}-\overline{\mathrm{PDR}}^{\,\text{attack}}.
\label{eq:potency}
\end{equation}
\end{linenomath}
\hfill\break
Mean speed tracks the effect on driving through time. It averages the instantaneous speed over the moving vehicles $\mathcal{V}(t)$ present at time $t$,
\hfill\break
\begin{linenomath}
\begin{equation}
\bar v(t)=\frac{1}{|\mathcal{V}(t)|}\sum_{i\in\mathcal{V}(t)}v_i(t).
\label{eq:meanspeed}
\end{equation}
\end{linenomath}
\hfill\break
Hard braking completes the picture, normalized per thousand vehicle-time samples so that runs of different length and occupancy compare directly. A hard-braking event fires when a vehicle's deceleration reaches $-4.0$\,m/s$^2$, and it rearms only after recovery, so a prolonged interaction counts once rather than at every time step. We also record a safety conflict when a following vehicle's time-to-collision falls to $3.0$\,s or below, on the same episode basis, and report below why that count is a poor discriminator here \citep{wang_review_2021}. We report each metric for the analytical baseline and the distilled model, under normal traffic and under the flooding condition of Equations~\eqref{eq:qa}--\eqref{eq:obu}.

\subsection{Scenario and Configuration}
The study uses two SUMO road networks. A single signalized intersection, spanning roughly $309\times181$\,m, supplies the fitting data; a two-intersection urban corridor of comparable extent, $218\times236$\,m, serves as the out-of-sample geometry. Each network carries one signalized junction per intersection. Demand holds constant while the connected share varies across six levels, 1, 5, 25, 50, 75, and 100 percent, so penetration alters the transmitter population without altering total traffic. Every connected vehicle broadcasts a 256-byte cooperative awareness message at 10\,Hz, matching the 100\,ms reservation interval.

Connected vehicles follow SUMO's cooperative adaptive cruise control model at a $0.6$\,s headway and a $1.0$\,m minimum gap, degrading to its adaptive cruise control model at $1.4$\,s and $2.0$\,m when the fail-safe of the next paragraph fires; both are the SUMO implementations \citep{lopez_microscopic_2018}, whose control laws derive from Milan\'es and Shladover \citep{MILANES2014285}. Unconnected vehicles and bicycles keep the default intelligent driver model throughout, so penetration changes the driving model only for vehicles that carry a radio. SUMO's own communication override for the cooperative model is left disabled, so message loss reaches driving through the reception model under test and through no other path.

The fail-safe has two responses, both keyed to message staleness. When a leader's awareness messages stop arriving for longer than $0.5$\,s, five consecutive losses at 10\,Hz, the follower drops from the cooperative to the adaptive controller, which widens its headway. When signal-phase messages go stale by the same margin within $35$\,m of a stop line, the vehicle decelerates to a halt at the line at $3.0$\,m/s$^2$, bringing itself to a minimal risk condition in the sense of \citet{sae_j3016_2021}. Both responses reverse on the next message that arrives, and both are active in every treatment arm of the trajectory experiments, so no arm is advantaged by a degradation the others cannot express.

Table~\ref{tab:scenario} lists the radio and resource-pool configuration. Both simulators read the same values, and the selection-window bounds $T_1$ and $T_2$ take the ns-3 defaults.

Each configuration runs five seeds. A repetition starts both simulators from the same demand realization: the SUMO seed follows the OMNeT run number, and that value also drives the trajectory export ns-3 replays. Runs last 200\,s at penetrations up to 50 percent and 60\,s at 75 and 100 percent, where vehicle counts make longer runs costly. The shorter horizon applies on the ns-3 side as well, which keeps every per-penetration comparison matched.

Five seeds place the treatments apart without resolving small differences within one. Per-seed delivery for a single configuration carries a standard deviation of up to $0.05$ across penetration levels and $0.02$ at the highest, putting the standard error of a five-seed mean near $0.02$. We therefore treat separations of that order as indistinguishable. The gaps between treatments exceed it by more than an order of magnitude.

Because traffic accumulates for less time, in-range density at 75 percent penetration does not exceed that at 50 percent. The mean in-range transmitter count measured in ns-3 5G-LENA rises from 8.5 at 5 percent to 58.2 at 50 percent, then reads 35.9 at 75 percent and 60.9 at 100 percent. The horizon therefore pairs 75 percent with 25 percent, and 100 percent with 50 percent, at nearly equal density. Delivery declines monotonically in measured density but not in penetration, so we report results against both.

\begin{table}[!ht]
\caption{Radio and Resource-Pool Configuration, Common to ns-3 5G-LENA and the Cascade}\label{tab:scenario}
\begin{center}
\small
\begin{tabular}{|l|l|l|}
\hline
\textbf{Parameter} & \textbf{Value} & \textbf{Note}\\
\hline
Carrier frequency $f_c$ & 5.9 GHz & ITS band \\
\hline
Channel bandwidth & 10 MHz & 5 subchannels, 10 resource blocks each \\
\hline
Numerology / subcarrier spacing & 0 / 15 kHz & 1 ms slot \\
\hline
Transmit power $P_{\mathrm{tx}}$ & 23 dBm & \\
\hline
Noise figure & 7 dB & noise floor $N_0=-94.28$ dBm \\
\hline
Modulation and coding scheme & 14 & 16-QAM, code rate $553/1024$ \\
\hline
Reservation interval $T_{RRI}$ & 100 ms & 256 B message at 10 Hz \\
\hline
Selection window $[T_1,T_2]$ & $[2,33]$ slots & TS 38.214 §8.1.4 \\
\hline
Sensing window $T_0$ / RSRP threshold & 100 ms / $-110$ dBm & sensing enabled \\
\hline
Max transmissions per reservation & 3 & HARQ enabled; 2.5 measured mean \\
\hline
Resource keeping $p_k$ / counter $R_c$ & $0$ / $U(5,15)$ & no persistent reservation \\
\hline
Channel model & TR 37.885 V2V-Urban & line-of-sight state per link \\
\hline
Communication range & 300 m & in-range threshold for all metrics \\
\hline
\end{tabular}
\end{center}
\end{table}

The ns-3 reference implements no application-layer security, so cryptographic delay is zero and receiver processing capacity is unbounded. Veins matches both, which leaves the channel model as the only difference between treatments. The overload stage of Equation~\eqref{eq:obu} is therefore inactive outside the adversarial runs, where the values of Table~\ref{tab:coeffs} reinstate it.

\section{Results and Discussion}
Results proceed from delivery, to the mechanisms producing it, to transfer across geometry, and then to the driving and adversarial outcomes. Three treatments run throughout, carrying the labels used in every figure and table: Plain, the channel the stack ships with; Analytical M3, the combined analytical reference of Equation~\eqref{eq:m3}; and NS3Learn, the distilled model. The same drop hooks inside OMNeT++ measure all three, so the comparison never depends on how a model derives its terms internally.

\subsection{Delivery Under Ordinary Traffic}
\begin{figure}[!ht]
  \centering
  \includegraphics[width=1\textwidth]{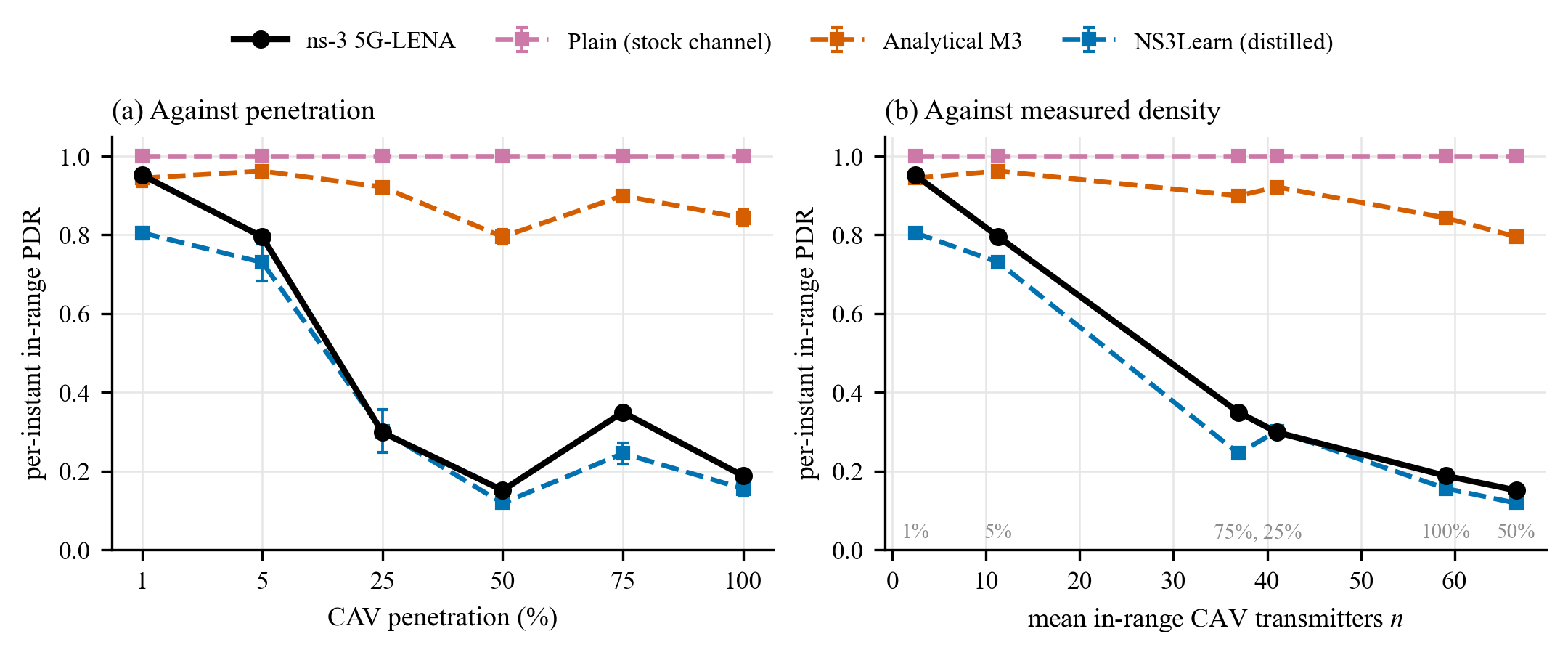}
  \caption{Per-instant in-range delivery against ns-3 5G-LENA, by penetration (a) and by measured density (b). Error bars are the standard deviation over five seeds. Labels in (b) group the penetration levels that share a density.}\label{fig:delivery}
\end{figure}

ns-3 5G-LENA establishes what a faithful Mode-2 channel does as vehicles accumulate. Per-instant in-range delivery falls from $0.952$ at 1 percent penetration to $0.188$ at 100 percent, a fall of more than four fifths driven by contention rather than distance, since the propagation environment is unchanged throughout. Figure~\ref{fig:delivery} places every treatment against that reference.

Plain, the unmodified channel, does not move. It is INET's \texttt{Ieee80211ScalarRadio} with the analytical channel bypassed, and it delivers $1.000$ at every penetration level. A scalar radio resolves a link budget without an interference model, so within range every message arrives and nothing in the configuration responds to a growing transmitter population. Its mean absolute deviation is $0.545$, rising to $0.812$ at the densest condition. A safety study run on this channel assumes every in-range neighbor hears every message, where ns-3 5G-LENA delivers under a fifth of them.

The combined analytical reference moves in the right direction and stops far short. Delivery declines from $0.944$ to $0.843$, a span of $0.10$ against ns-3 5G-LENA's $0.76$, giving a mean absolute deviation of $0.441$ and a worst case of $0.655$ at 100 percent penetration. Joining a published collision model to a published propagation model, each evaluated on ns-3 5G-LENA's own channel and link configuration, recovers roughly one eighth of the density response. The reference is calibrated for a regime whose collision probability never approaches what ns-3 5G-LENA measures here.

NS3Learn tracks ns-3 5G-LENA across the full range at a mean absolute deviation of $0.064$. That figure is a goodness of fit, since NS3Learn is fitted to ns-3 5G-LENA labels and then scored against them, whereas the deviations of the two references are the errors of models that never saw the target. We report the three side by side to measure how far a model built without contention can track a channel governed by it, rather than as a competitive ranking.

The residual is close to one-sided. NS3Learn under-delivers at five of the six penetration levels and on the corridor, giving a signed mean bias of $-0.063$ against a mean absolute deviation of $0.064$. The single exception is 25 percent, where it over-delivers by $0.004$, an order of magnitude inside the per-seed spread at that level. Agreement is closest in the dense conditions: $0.302$ against $0.298$ at 25 percent, $0.119$ against $0.151$ at 50 percent, and $0.156$ against $0.188$ at 100 percent.

The largest error sits at the sparsest condition, where NS3Learn returns $0.805$ against $0.952$ and overstates loss by $0.147$. That condition carries a mean of $2.4$ in-range transmitters, below the density range the collision and half-duplex stages were fitted over, and the logistic forms of Equations~\eqref{eq:hd} and~\eqref{eq:coll0} retain a floor there that ns-3 5G-LENA does not exhibit: evaluated at $n=0$ they still return $0.082$ and $0.065$. Constraining those stages to pass through the origin removes it, and is the first refinement we would make. The bias runs pessimistic almost throughout, which matters because the downstream conclusions in this paper are also pessimistic; a reader should weigh them knowing the model errs in that direction at every point we measured except 25 percent, where it is effectively unbiased.

Panel (b) resolves an apparent anomaly in the penetration axis. Delivery does not fall monotonically in penetration: the 75 percent condition recovers to $0.349$ from $0.151$ at 50 percent. In measured density it falls monotonically. The shortened horizon at the two highest penetrations pairs 75 percent with 25 percent at $36.9$ and $41.1$ in-range transmitters, and 100 percent with 50 percent at $59.1$ and $66.6$. ns-3 5G-LENA delivers a comparable value within each pair, $0.349$ against $0.298$ and $0.188$ against $0.151$, the residual gap tracking the residual density gap, and NS3Learn reproduces both. Density, not connected share, sets Mode-2 delivery, and a model responding to density inherits that behavior automatically.

\subsection{Where Each Model Places Its Loss}
\begin{figure}[!ht]
  \centering
  \includegraphics[width=1\textwidth]{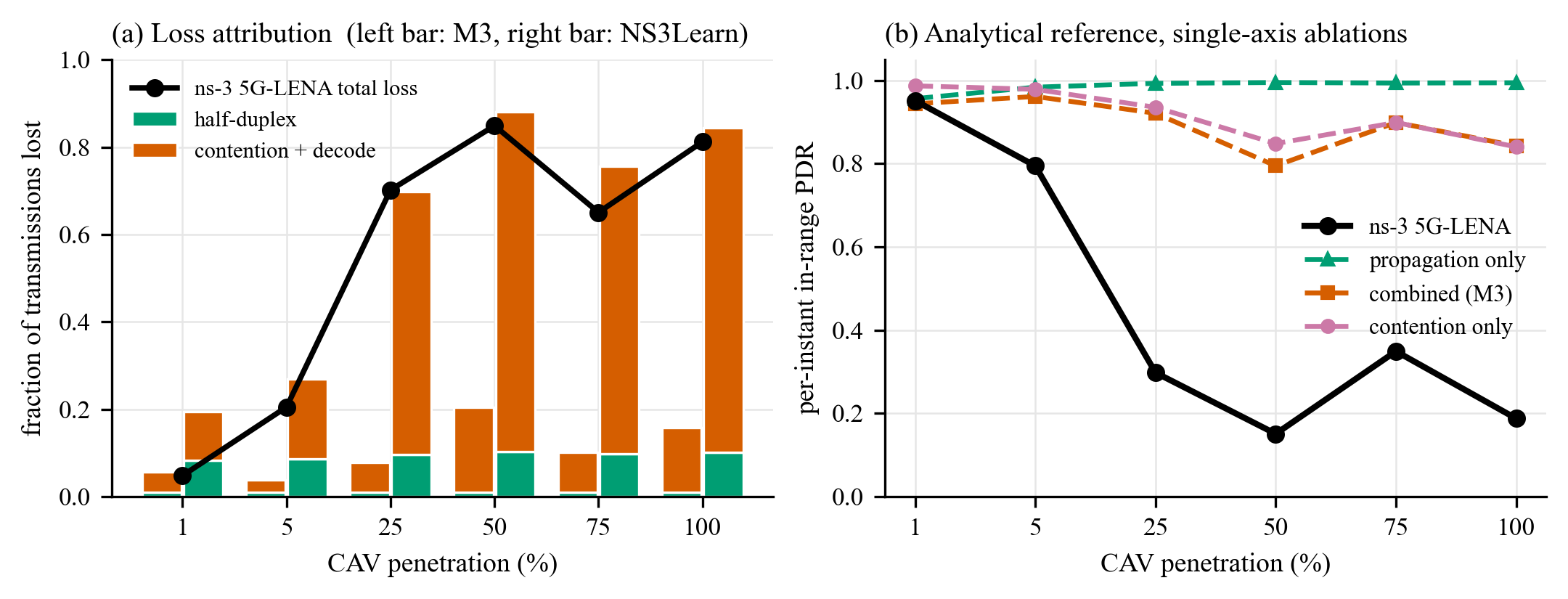}
  \caption{Loss attribution against ns-3 5G-LENA's total loss (a), and single-axis ablations of the analytical reference (b). In (b) the contention-only and combined curves nearly coincide, meeting at 75 and 100 percent, while propagation alone tracks the untouched stack.}\label{fig:attribution}
\end{figure}

Matching a delivery level does not establish that a model matches the mechanism, so we label each discarded message by stage that discarded it. Figure~\ref{fig:attribution}(a) compares the two attributions against ns-3 5G-LENA's total loss.

The half-duplex term separates them. The analytical reference sets it to the share of time a vehicle spends transmitting instead of listening, $P_{\mathrm{HD}}=t_s/T_{RRI}$, a ratio of durations that cannot change as vehicles are added. ns-3 5G-LENA blocks $0.091$ of receptions at the sparsest condition and $0.105$ at the densest, and the fitted stage of Equation~\eqref{eq:hd} follows it from $0.083$ to $0.103$ because it reads the neighbor count. What separates the two is the dependence, not the level. Granting the reference the same mean blocking fraction the fitted stage produces, $0.100$ against $0.102$ and at the top of the range ns-3 5G-LENA measures, still leaves its mean absolute deviation at $0.389$ against $0.064$, six times larger at matched average blocking. Smaller constants do worse: $0.441$ at the single-transmission value the reference implementation specifies, $0.433$ at the $0.025$ implied by the $2.5$ transmissions each reservation carries on average, and $0.423$ at $0.040$. No constant closes the gap, because the closed form omits a dependence rather than a value. What keeps the ns-3 5G-LENA receiver blocked so much longer than its own transmissions last is a property of that implementation, and NS3Learn reproduces it without explaining it.

Contention accounts for the rest of the divergence. Loss in ns-3 5G-LENA reaches $0.812$ at the densest condition, of which NS3Learn attributes $0.742$ to contention and decode and $0.102$ to half-duplex, summing to slightly above the observed level. The analytical reference charges $0.147$ to contention there, falling short by a factor of five. The shortfall is definitional as much as numerical. Our labels mark a transmission as collided when another overlaps its resource blocks in the same slot, counted across the whole network rather than within communication range. A mean of seven transmissions share each slot at the densest conditions, against five subchannels. Equation~\eqref{eq:pcol} resolves collisions among the in-range population alone, over a $124$-resource selection window. The two quantities therefore differ whatever the merits of either, and we report the gap as a discrepancy to resolve rather than as evidence that the published model understates contention. Capture makes the difference tractable: ns-3 5G-LENA's high collision rate does not translate into equally high loss, because the stronger of two overlapping signals frequently survives. A model without a capture branch must either understate collisions or overstate loss, and the reference takes the first path.

Panel (b) isolates each analytical factor. Propagation alone is nearly indistinguishable from the untouched stack, at $0.531$ mean absolute deviation against Plain's $0.545$, and its delivery rises slightly with penetration because denser traffic shortens the mean in-range link. Contention alone reaches $0.459$, and adding propagation improves the combination to $0.441$. Contention therefore carries nearly all of the reference's explanatory power, with propagation contributing about $0.02$ across a range in which ns-3 5G-LENA moves by $0.76$. For NR Mode-2 in urban traffic, resolving propagation without resolving medium access performs about as well as no channel model.

\subsection{Transfer to an Unseen Geometry}
The corridor test uses the intersection coefficients unchanged, with no refitting, against an ns-3 5G-LENA reference run on the same geometry. On the corridor, ns-3 5G-LENA delivers $0.390$ at 50 percent penetration, with a standard deviation of $0.002$ across three seeds and a mean of $34.6$ in-range transmitters.

The corridor at 50 percent penetration sits nearest the intersection's 75 percent density, $34.6$ against $36.9$, and the two runs deliver $0.390$ and $0.349$. Changing road layout, junction count, and signal timing shifts delivery by $0.04$ once density is held approximately level, which identifies density as the variable that transfers.

Frozen, NS3Learn returns $0.266$ on the corridor, a deviation of $0.124$, against $0.105$ at the matched in-sample density. The deviation grows by about 20 percent when the geometry changes, and the residual keeps its direction, the same conservative overstatement of loss seen at low density. The analytical reference returns $0.916$, a deviation of $0.526$. Transfer therefore costs NS3Learn some accuracy and leaves it more than four times closer to the out-of-sample reference than the reference, with no parameter re-estimated.

\subsection{Consequences for Driving}
\begin{figure}[!ht]
  \centering
  \includegraphics[width=0.90\textwidth]{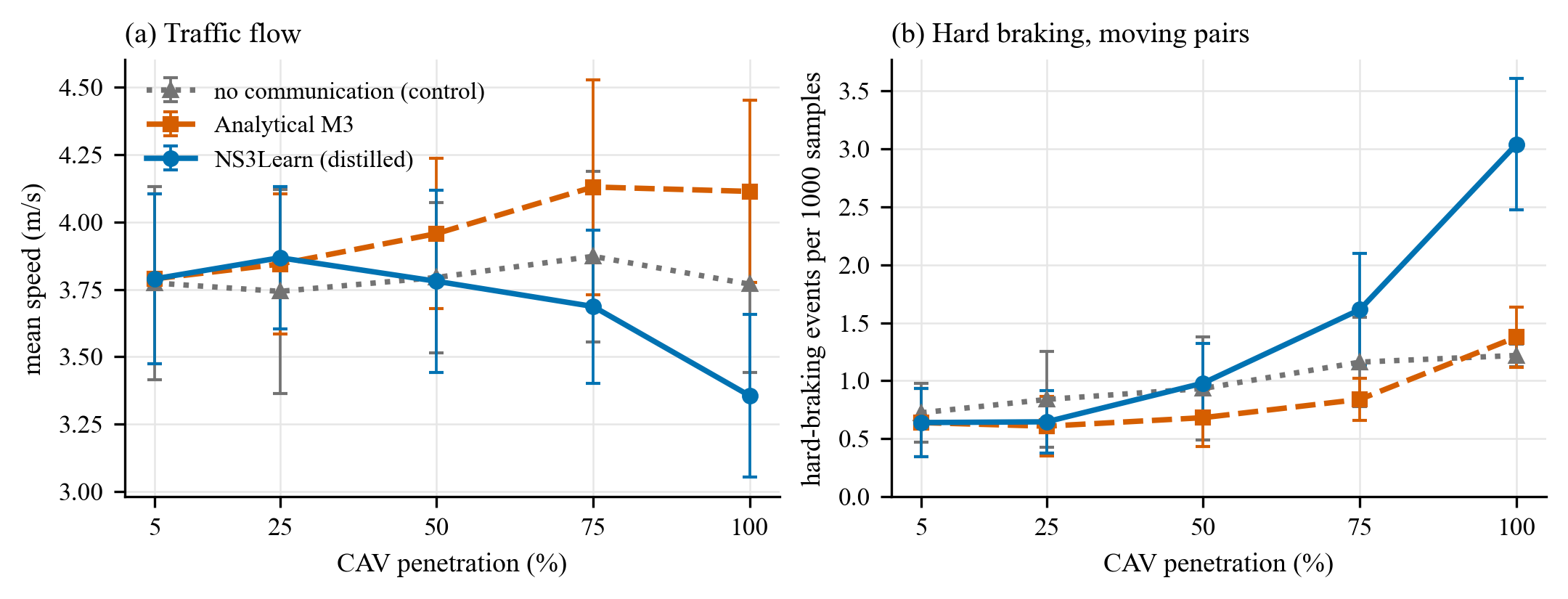}
  \caption{Traffic outcomes under the two reception models and a control with the application disabled, five seeds, fail-safe active in both treatment arms. Mean speed (a) moves in opposite directions relative to the control. Hard braking (b) counts moving pairs only; the analytical reference tracks the control closely while NS3Learn separates from both above 50 percent penetration.}\label{fig:trajectory}
\end{figure}

Reception models earn their cost only if they change what a study concludes. The network-side runs enable the message-driven fail-safe in every arm, so a vehicle acts on what it receives and no arm can avoid a degradation the others express, and swapping the reception model while holding traffic, demand, seeds, and car-following model fixed isolates the effect. Figure~\ref{fig:trajectory} shows the two models moving in opposite directions.

Under the analytical reference, mean speed rises from $3.79$ to $4.11$\,m/s as penetration grows, which credits connectivity with improved flow. Under NS3Learn the same scenario yields mean speed falling from $3.79$ to $3.36$\,m/s. The sign of the trend is the first divergence, and it comes from a direct measurement that needs no interpretation.

Under NS3Learn at full penetration, $89$ percent of recorded conflicts involve a leader below $0.5$\,m/s, and the following vehicle's own median speed is $0.11$\,m/s: these are vehicles inching toward a stationary queue, not vehicles closing at speed. Message loss arms the fail-safe, which stops vehicles and forms queues. A metric that divides gap by closing speed then fires on every pair in the queue. We therefore report conflicts twice. Counted over all samples, exposure rises from $2.82$ to $20.55$ per thousand and reaches $1.9$ times the reference at full penetration. Restricted to pairs where the follower exceeds $2$\,m/s and the leader exceeds $1$\,m/s, which retains $27$ percent of samples at full penetration, it rises from $2.25$ to $5.40$ and reaches $1.4$ times the reference. The gated ratio grows with penetration, from $0.99$ at 5 percent to $1.42$ at 100, though not monotonically. The divergence is therefore real, and its raw magnitude is inflated by about a third by queue formation.

Hard braking responds to the gate in the opposite direction. It separates by $1.2$ times before gating and $2.2$ times after, at $1.38$ against $3.04$ events per thousand moving-pair samples, because braking under NS3Learn concentrates among vehicles still in motion while the analytical arm accumulates its events in slow traffic that the gate removes. Gating therefore strengthens rather than deflates this comparison, and Figure~\ref{fig:trajectory}(b) shows the separation opening from 50 percent penetration onward.

A third arm gives the outcome connectivity is measured against: the same scenario with the application disabled, so that no vehicle acts on any message. At full penetration that baseline reaches $3.77$\,m/s, with $4.58$ gated conflicts and $1.22$ hard-braking events per thousand samples. The analytical reference places connectivity above the baseline on flow and on exposure, at $4.11$\,m/s and $3.81$ conflicts, and marginally below it on hard braking at $1.38$ events. NS3Learn places it below the baseline on every count, at $3.36$\,m/s, $5.40$ conflicts and $3.04$ events. From identical traffic inputs, differing only in which messages arrive, the two channel models therefore disagree about whether this cooperative application is worth deploying at all. The baseline also bounds what the result covers: a fail-safe that leaves traffic worse than no connectivity is a statement about that design as much as about the radio, and what we show is that the channel model decides which of the two verdicts a study reports.

Delivery explains the divergence. A channel that keeps delivering under load lets cooperative behavior operate as designed. A channel that degrades as neighbors accumulate leaves vehicles acting on stale information when the intersection is busiest. The reference still delivers $0.843$ at full penetration and reports the optimistic outcome. NS3Learn delivers $0.156$, close to ns-3 5G-LENA's $0.188$, and reports the pessimistic one. A study adopting the reference would conclude that dense deployment is safe. The reference implementation contradicts that conclusion.

\subsection{Response to an Adversarial Flood}
A flooding adversary tests whether a model responds to load that is concentrated rather than distributed. The attacker broadcasts at $1000$\,pps against a $10$\,Hz background, and ns-3 5G-LENA's own flood traces give the reference response. Receiver overload is active in these runs and dominates the delivery ratio, which makes the change in total delivery a poor discriminator: it is largest for the models delivering most at baseline, so it rewards under-delivery. Table~\ref{tab:flood} therefore reports the channel-side response, the change in loss charged to contention and decode. That quantity is what the adversary acts on, and it compares directly against ns-3 5G-LENA.

ns-3 5G-LENA's collision probability rises by $0.393$ at low density, by $0.089$ at the middle density, and by only $0.014$ at the highest, where baseline collision already stands at $0.96$ and leaves an adversary little room to add. NS3Learn responds where that response is large, charging an extra $0.176$ to channel-side loss at the sparsest condition, three times its seed-to-seed spread. It recovers a little under half of ns-3 5G-LENA's rise rather than matching it. At the middle density it charges $0.046$ against a spread of $0.014$, so the response resolves there as well. At the highest density ns-3 5G-LENA moves by about one hundredth, and five seeds do not separate the NS3Learn estimate from zero, so the comparison there is inconclusive.

The two references carry no such response. The analytical reference moves the other way, by $-0.003$ on average and resolvably so at low density. Its collision term reads a neighbor count that excludes an active flooder, so introducing an attacker slightly thins the contending population it sees. Plain changes by exactly zero. Neither reference has a term any flood rate could act on. The analytical collision probability is a function of the in-range vehicle count, so an adversary transmitting a hundred times more often than its neighbors still enters as one vehicle. Plain carries no contention term at all. No parameter choice changes this, because the mechanism is absent rather than mis-tuned.

Protocol-aware attacks on SB-SPS resource selection draw attention because they exploit the scheduler \citep{twardokus_toward_2023,ashik_analyzing_2026}. A model whose channel term cannot register a flooder reports such an attack as harmless at the channel and charges its damage entirely to receiver processing. That sets the boundary of what each representation supports: NS3Learn carries a mechanism the flood rate enters, and the alternatives carry none. Sizing that response across attacker rates and populations needs a wider sweep than the single rate reported here.

\begin{table}[!ht]
\caption{Channel-Side Response to a 1000\,pps Flooding Adversary}\label{tab:flood}
\begin{center}
\small
\begin{tabular}{|l|c|c|c|c|c|}
\hline
& \multicolumn{3}{c|}{\textbf{ns-3 5G-LENA collision prob.}} & \multicolumn{2}{c|}{\textbf{Change in channel-side loss}}\\
\cline{2-6}
\textbf{Density $n$} & \textbf{Baseline} & \textbf{Attacked} & \textbf{Change} & \textbf{Analytical M3} & \textbf{NS3Learn}\\
\hline
$8.8$   & 0.196 & 0.589 & $+0.393$ & $-0.014\pm0.004$ & $+0.176\pm0.056$\\
\hline
$33.7$  & 0.799 & 0.888 & $+0.089$ & $-0.001\pm0.002$ & $+0.046\pm0.014$\\
\hline
$65.9$  & 0.957 & 0.971 & $+0.014$ & $+0.006\pm0.010$ & $+0.007\pm0.007$\\
\hline
Mean    & --    & --    & $+0.165$ & $-0.003$ & $+0.076$\\
\hline
\end{tabular}
\end{center}
\footnotesize Plain changes by $0.000$ at every density and is omitted. The ns-3 5G-LENA baseline column is measured on the symmetric $10$\,Hz sweep, pooling $1.82$M labeled transmissions, and the attacked column on the flood runs, pooling $1.06$M. Cascade and reference columns are means over five matched baseline and attacked seeds, $\pm$ one standard deviation across seeds; the two sparser conditions separate from zero. Receiver-overload loss is excluded here and discussed in the text.
\end{table}

\section{Conclusions}
Medium access governs NR PC5 Mode-2 delivery in urban traffic, and the propagation-oriented channels the OMNeT++, Veins, and SUMO stack inherited from its 802.11p origins cannot represent it. Measured against 3GPP-calibrated ns-3 5G-LENA, whose delivery falls from $0.952$ to $0.188$ as connected vehicles accumulate, the unmodified channel does not move, and a combined analytical reference built from two published Mode-2 models recovers about one eighth of that decline. Distilling ns-3 5G-LENA into NS3Learn, a five-stage logistic cascade, closes the gap to a mean absolute deviation of $0.064$. The result is closed-form, evaluated per message, transparent in its coefficients, and free of any new protocol module.

Three findings extend beyond this scenario. Attribution matters as much as accuracy. A slot-fraction half-duplex term cannot grow with density, and matching its average to the measured blocking fraction still leaves it six times further from the reference than a term that reads the neighbor count, so two models can reach one delivery figure through mechanisms that diverge under any change of conditions. Density rather than penetration is the governing variable, shown by matched-density pairs that deliver alike across penetration levels. Capture is a requirement rather than a refinement, because ns-3 5G-LENA still delivers a fifth of its messages where its collision rate exceeds $0.9$.

Holding traffic, demand, seeds, and car-following model fixed and changing only the reception model more than doubles hard braking and moves the outcome across the no-communication control: the analytical reference places connectivity above that control on flow and exposure, and NS3Learn places it below on every metric. Under an adversarial flood the difference is one of kind rather than degree. NS3Learn carries a channel-side mechanism the flood rate enters and registers $0.176$ of additional loss where ns-3 5G-LENA's own response is largest; the analytical reference and the unmodified channel carry no such term at any flood rate. Results on communication-dependent safety applications, cooperative automation, and sidelink attack studies are therefore contingent on a modeling choice rarely stated as one.

This fidelity is now available without leaving the stack. Distillation costs one offline labeling and fitting campaign, and the fitted numbers then travel as a plain-text file the simulation reads when a run starts. Moving to another numerology, message rate, or resource pool means repeating that campaign rather than porting a protocol implementation. The recipe applies wherever a mature reference implementation exists in one simulator and the experiment must run in another. We release NS3Learn, its coefficients, and the distillation and analysis code, so that researchers and state ITS teams evaluating connected-vehicle safety on this stack can substitute a contention-aware channel for a propagation-only one and report results that describe the medium their vehicles share.

Four limits bound the claims. Distillation reproduces its source, including whatever that source has wrong. We call ns-3 5G-LENA a reference implementation rather than ground truth for that reason, and the half-duplex residual is a concrete instance: a measured blocking fraction of $0.091$ to $0.105$ against a slot fraction of $0.010$, or $0.025$ once the average transmission count is folded in, may reflect how the sidelink receiver is implemented rather than how a radio behaves. Nothing in our method detects such an error, and a model fitted to it carries the error forward with the realism.

Scope is narrow. We fit NS3Learn at one urban intersection under one channel model, numerology, and message rate, and demonstrate transfer to a second signalized urban junction rather than to a highway or another band. Both layouts concentrate vehicles at controlled stops, so the transfer probes layout, scale, and signal timing, not a contrast between clustered and dispersed traffic.

Accuracy carries a systematic bias whose cause is identified rather than open. The residual is one-sided at five of the six operating points and runs conservative, reporting more loss than the reference rather than less. Its mechanism is the logistic floor described above, a property of the functional form rather than of the data, and it binds only below the density range the stages were fitted over; constraining both to pass through the origin removes it. Two questions stay open. Our collision label counts an overlap anywhere in the network while Equation~\eqref{eq:pcol} resolves collisions in range alone, so the two are not measured over the same set. And the reference implementation's collision probability rises with attacker rate, by $0.339$ at 200 packets per second and $0.393$ at 1000 at the sparsest density, while we evaluate the surrogate at the higher rate alone. Sweeping attacker rate and population would settle both.

\section{ACKNOWLEDGMENT}
The authors acknowledge the use of Anthropic (2026), Opus 4.8 for controlled python code generation, code audit, compilation of relevant sources, grammar and writing error audits only. 

\section{AUTHOR CONTRIBUTIONS}
The authors confirm contribution to the paper as follows: study conception and design: \textbf{\textit{R. Bello, J. Mwakalonge, J. Sahoo}}; data collection: \textbf{\textit{R. Bello, A. Mukwaya}}; analysis and interpretation of results: \textbf{\textit{R. Bello, G. Comert, V. Bendigeri, V. Vaidyan}}; draft manuscript preparation: \textit{\textbf{R. Bello, A. Mukwaya, A. Dontoh}}. All authors reviewed the results and approved the final version of the manuscript.

\section{DECLARATION OF CONFLICTING INTERESTS}
The authors declared no potential conflicts of interest with respect to the research, authorship, and/or publication of this article.

\section{FUNDING}
This research was partly funded by the U.S. Department of Education through the HBCU Master's Program Grant (Grant No. P120A210048); the U.S. Department of Transportation's University Transportation Centers Program grant, administered by the Transportation Program at South Carolina State University (SCSU); and the National Science Foundation (NSF) under Grant Nos. 2131080, 2242812, 2200457, 2234920, and 2305470.

\clearpage
\nolinenumbers
\printbibliography[title={REFERENCES}]

\end{document}